\documentclass[9pt, twocolumn]{extarticle}
\usepackage{multicol}
\usepackage[]{graphicx}
\usepackage[]{xcolor}
\usepackage{alltt}
\usepackage[T1]{fontenc}
\usepackage[utf8]{inputenc}
\usepackage{multirow}
\usepackage{gensymb}
\usepackage{subfigure}
\usepackage{amsmath}
\usepackage{cite}
\usepackage{array}
\usepackage[]{authblk}
\usepackage{threeparttable}
\usepackage{textcomp}
\usepackage{soul,xcolor}
\usepackage{caption}
\usepackage{pdfpages}

\usepackage{xr}
\makeatletter
\newcommand*{\addFileDependency}[1]{
  \typeout{(#1)}
  \@addtofilelist{#1}
  \IfFileExists{#1}{}{\typeout{No file #1.}}
}
\makeatother

\usepackage[colorlinks,
			linkcolor=blue,
			anchorcolor=green,
			citecolor=blue,
			urlcolor=blue
			]{hyperref}
\usepackage[nameinlink,noabbrev]{cleveref} 

\usepackage{fancyhdr}
\usepackage[a4paper,top=60pt,bottom=60pt,left=50pt,right=50pt]{geometry}

\usepackage{textcomp}

\def\keywords{\vspace{.3em}
{\textit{Keywords}:\,\relax%
}}
\def\endkeywords{\par}

\usepackage{float}
\floatstyle{plaintop}
\restylefloat{table}

\author[1,2]{Xiang Xu}
\author[2,*]{Xi Zhang}
\author[3,4]{Andrei Ruban}
\author[1]{Siegfried Schmauder}
\author[2]{Blazej Grabowski}
\affil[1]{Institute for Materials Testing, Materials Science and Strength of Materials, University of Stuttgart, Pfaffenwaldring 32, 70569 Stuttgart, Germany}
\affil[2]{Institute for Materials Science, University of Stuttgart, Pfaffenwaldring 55, 70569 Stuttgart, Germany}
\affil[3]{KTH Royal Institute of Technology, SE-100 44 Stockholm, Sweden}
\affil[4]{Materials Center Leoben Forschung GmbH, A-8700 Leoben, Austria}
\affil[*]{\rm Corresponding author: xi.zhang@imw.uni-stuttgart.de}

\title{Accurate complex-stacking-fault Gibbs energy in Ni$_3$Al at high temperatures}
\begin{document}
\date{}
\setstcolor{magenta}
\twocolumn[
\begin{@twocolumnfalse}
\maketitle
\section*{Abstract}

To gain a deeper insight into the anomalous yield behavior of Ni$_3$Al, it is essential to obtain temperature-dependent formation Gibbs energies of the relevant planar defects. Here, the Gibbs energy of the complex stacking fault (CSF) is evaluated using a recently proposed \textit{ab initio} framework ${[}$Acta~Materialia, \textbf{255} (2023) 118986${]}$, accounting for all thermal contributions---including anharmonicity and paramagnetism---up to the melting point. The CSF energy shows a moderate decrease from 300\,K to about 1200\,K, followed by a stronger drop. 
We demonstrate the necessity to carefully consider the individual thermal excitations.
\textcolor{black}{
We also propose a way to analyze the origin of the significant anharmonic contribution to the CSF energy through atomic pair distributions at the CSF plane.
With the newly available high-temperature CSF data, an increasing energy barrier for the cross-slip process in Ni$_3$Al with increasing temperature is unveiled, necessitating the refinement of existing analytical models.}
\\
\keywords
Yield Stress Anomaly; Complex Stacking Fault; Ab-initio Calculations; Longitudinal Spin Fluctuations.
\endkeywords

\vspace{1cm}
\end{@twocolumnfalse}

]


The yield strength of metallic materials typically decreases with increasing temperature due to thermally activated dislocation motion. However, several L1$_2$-ordered compounds, such as Ni$_3$Al and Ni$_3$Ga, exhibit an abnormal increase in their yield strength from 77\,K to about 1100\,K, which is known as the yield stress anomaly~(YSA)~\cite{takeuchi1973temperature,mishima1986mechanical,golberg1998effect}. Such an anomalous yield behavior makes these compounds the key components for high-temperature applications such as turbine blades~\cite{furrer1999ni}. 

The YSA of L1$_2$ Ni$_3$Al originates from the core structure of the $a\left<110\right>$ superdislocation dissociating into two  $a/2\left<110\right>$ superpartials bound by antiphase boundaries~(APBs).
Each superpartial can further split into two Shockley partials,  generating a complex stacking fault~(CSF) in between.
At elevated temperatures, the cross-slip of screw superpartials from an easy-glide octahedral plane to a less favorable cubic plane forms Kear-Wilsdorf (KW) locks that strengthen the material~\cite{sun1988tem,karnthaler1996influence,kruml2002dislocation}.
To explain the cross-slip behavior and thus the YSA, various models~\cite{paidar1984theory,yoo1987stability,hirsch1992model,caillard1997role,choi2007modelling,demura2007athermal} have been proposed based on the formation energies of the three involved planar defects (CSF and two types of APBs). 
It has been generally concluded that it is the CSF energy that tunes the cross-slip process~\cite{baluc1991tem,karnthaler1996influence} and thus the peak  temperature of the YSA~\cite{kruml2002dislocation}.

Despite the crucial role in interpreting the YSA, the temperature dependence of the CSF energy of Ni$_3$Al has not been rigorously evaluated either experimentally or theoretically.
Experimentally, the CSF energy is determined by measurement of the dissociation distance between Shockley partials and subsequent application of anisotropic elasticity theory. 
The measurements encounter significant challenges and uncertainties due to the short dissociation distances (1--2\,nm) of edge dislocations~\cite{hemker1993measurements,karnthaler1996influence}. 
Theoretically, accurate temperature-dependent CSF energies from \textit{ab initio} are
missing since calculations performed so far have been limited to 0\,K~\cite{shang2020unveiling,yu2012effect,wen2012first,tan2019dislocation} or low-temperature approximations~\cite{liu2015stacking}, with little consideration given to the impact of magnetism which has been shown to be strong for the APB energies~\cite{xu2023APBs}. 

In the present work, we aim at an accurate prediction of the temperature-dependent CSF Gibbs energy in Ni$_3$Al by utilizing a recently proposed \textit{ab initio} framework~\cite{xu2023APBs} that incorporates explicit lattice vibrations, electronic
excitations, and the impact of magnetic excitations up to the melting temperature. Special emphasis is placed on the physical origin of the strong impact of anharmonicity and spin fluctuations. Based on the obtained accurate CSF energies, new insights into the role of the CSFs in YSA are discussed.

\begin{figure}[!b]
\centering
\includegraphics[scale=0.9]{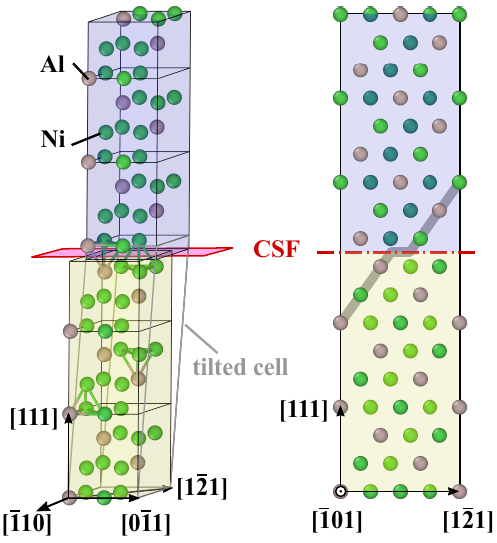}
\caption{Atomic model of the complex stacking fault.}
\label{fig:CSF}
\end{figure}

In the \textit{ab initio} framework based on the supercell approach, the pressure $P$ and temperature $T$ dependent CSF Gibbs energy $\gamma_\mathrm{CSF}$ is expressed as
\begin{equation}\label{eq_Gibbs_CSFE}
\gamma_\mathrm{CSF}(P,T)=\frac{G_{\mathrm{CSF}}(P,T)-G_{\mathrm{bulk}}(P,T)}{A_\mathrm{CSF}(P,T)}, 
\end{equation}
where $G_\mathrm{CSF}(P,T)$ and $G_\mathrm{bulk}(P,T)$ are the Gibbs energies of a supercell with the CSF and of the perfect bulk, respectively. 
Further, $A_\mathrm{CSF}$ represents the CSF area which can be derived from the lattice constant.
The utilized CSF model for nonmagnetic calculations is shown in Figure~\ref{fig:CSF}, i.e., 
a tilted supercell with nine atomic layers along the [111] direction.
For the magnetic calculations, a six-layer supercell is used. The selected supercell sizes ensure sufficient separation
between periodic
image CSFs introduced by the periodic boundary conditions.

It is convenient to calculate first the volume- and temperature-dependent Helmholtz energy $F(V,T)$ and then derive the Gibbs energy from a Legendre transformation. 
The Helmholtz energy is decomposed as 
\begin{equation}
\begin{aligned}\label{eq_Holm_contris}
    F(V,T) =\ & E_{\mathrm{0K}}(V)+ F^{\mathrm{qh}}(V,T) +F^{\mathrm{ah}}(V,T) \\
    & + F^{\mathrm{el}}(V,T) + F^{\mathrm{mag}}(V,T),
\end{aligned}
\end{equation}
where the terms on the right-hand-side denote the free energy contributions from 0~K, quasiharmonic (qh) and anharmonic (ah) thermal vibrations, and electronic (el) and magnetic (mag) excitations, respectively.

To account for the various free energy contributions, we employ state-of-the-art finite-temperature \textit{ab initio} approaches.
The finite displacement
approach is used to compute the quasiharmonic free energy $F^{\mathrm{qh}}$. The explicit anharmonic free energy $F^{\mathrm{ah}}$ is computed with the direct upsampling technique \cite{zhou2022thermodynamics,jung2023high} aided by machine-learning interatomic potentials, i.e., moment tensor potentials (MTPs)~\cite{shapeev2016moment,gubaev2019accelerating}, to achieve the required density-functional-theory (DFT) accuracy.
\textcolor{black}{An accurate MTP (RMSE: 0.31 meV/atom) trained on 2449 CSF structures sampled from 
molecular dynamics (MD) simulations (performed with VASP~\cite{kresse1993ab,kresse1994ab}) is used.}
The electronic free energy $F^{\mathrm{el}}$ including the coupling to thermal vibrations is computed in a second upsampling step utilizing finite temperature DFT~\cite{jung2023high,xu2023APBs}.
\textcolor{black}{The first four terms in Eq.~\eqref{eq_Holm_contris} correspond to spin-unpolarized calculations.}

The magnetic free energy $F^{\mathrm{mag}}$ is calculated based on the single-site mean-field approximation for the disordered local moment (DLM) paramagnetic (PM) state as implemented in the Lyngby version of the exact-muffin-tin orbital (EMTO) code~\cite{ruban2016atomic}. 
Within this approach, thermal magnetic excitations, particularly from longitudinal spin fluctuations (LSFs), are obtained by a self-consistent minimization of a generalized free energy expression,
\begin{equation}
\label{Fmag}
F^\mathrm{mag}(V,T) = \underset{\{m_i\}}{\min} \left[ E^\mathrm{mag}(\{m_i\}) - T S^\mathrm{mag}(\{m_i\}) \right],
\end{equation}
where $E^\mathrm{mag}(\{m_i\})$ is the  energy computed
for a DLM spin configuration with $m_i$ denoting the mean local magnetic moment at site $i$.
The magnetic entropy $S^\mathrm{mag}$ is calculated as~\cite{ruban2013impact}
\begin{equation}\label{eq_LSF}
    S^\mathrm{mag}(\{{m}_i\})= 3 \sum_i \ln(m_i).
\end{equation}
\textcolor{black}{This equation applies in the \emph{magnetic} high temperature limit when spin fluctuations are fully exited. Due to the low Curie temperature of Ni$_3$Al of 41.5\,K, the magnetic high-temperature limit can be justifiably assumed to apply at 300\,K and onward. This was verified earlier by Monte Carlo tests \cite{xu2023APBs}. Note that Eq.~\eqref{eq_LSF} is a classical model that yields relative changes of the LSF entropy, which can be negative when magnetic moments are less than one. This is not a problem if one compares energy differences at the same temperature as is the case for the CSF calculations. The usage of Eq.~\eqref{eq_LSF} is further supported by Ref.~\cite{khmelevskyi2018longitudinal} in which different forms of the magnetic entropy were tested.}

\textcolor{black}{
In our approach to the CSF energy, the coupling between magnetic excitations and thermal vibrations is not considered. The influence of atomic vibrations on spin fluctuations was shown to be negligible for the APBs in Ni$_3$Al~\cite{xu2023APBs}.
An explicit evaluation of the influence of magnetic excitations on atomic vibrations is not yet available for Ni$_3$Al. 
However, for fcc Ni (similar itinerant nature as Ni$_3$Al), it was shown that atomic forces and phonons are unaffected by magnetic excitations~\cite{kormann2016impact}. 
It seems therefore reasonable to assume that this coupling effect will be likewise small in Ni$_3$Al and that the CSF energy will not be affected.} Other technical
details related to the simulations can be found in Ref.~\cite{xu2023APBs}.

\begin{figure}[!b]
\centering
\includegraphics[scale=0.75]{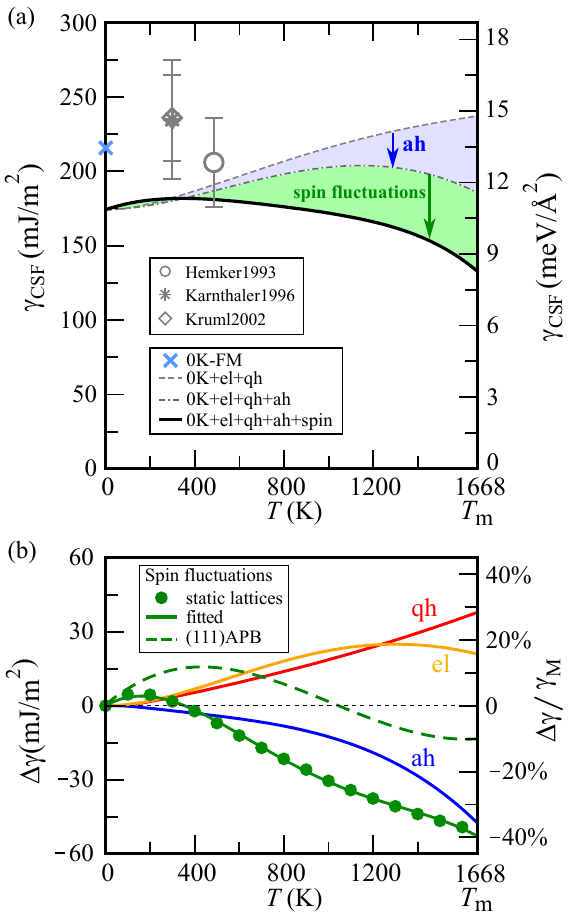}
\caption{(a) Temperature dependent Gibbs energy of the CSF. 
The symbols with error bars are the experimentally measured values, i.e., Hemker1993~\cite{hemker1993measurements}, Karnthaler1996~\cite{karnthaler1996influence}, Kruml2002~\cite{kruml2002dislocation}.
(b) The resolved contributions to the CSF energy ($\gamma_{\rm M}$: the CSF energy at the melting point).
The contribution 
\textcolor{black}{of spin fluctuations} is fitted with a third-order polynomial. 
The contribution 
\textcolor{black}{of spin fluctuations} for the (111)APB 
~\cite{xu2023APBs} was added for comparison.
The experimental melting temperature of Ni$_3$Al is $T_{\rm m}=1668$~K~\cite{dey2003physical}.
}
\label{fig:CSFE}
\end{figure}

\begin{figure*}[!ht]
\centering
\includegraphics[width=0.95\textwidth]{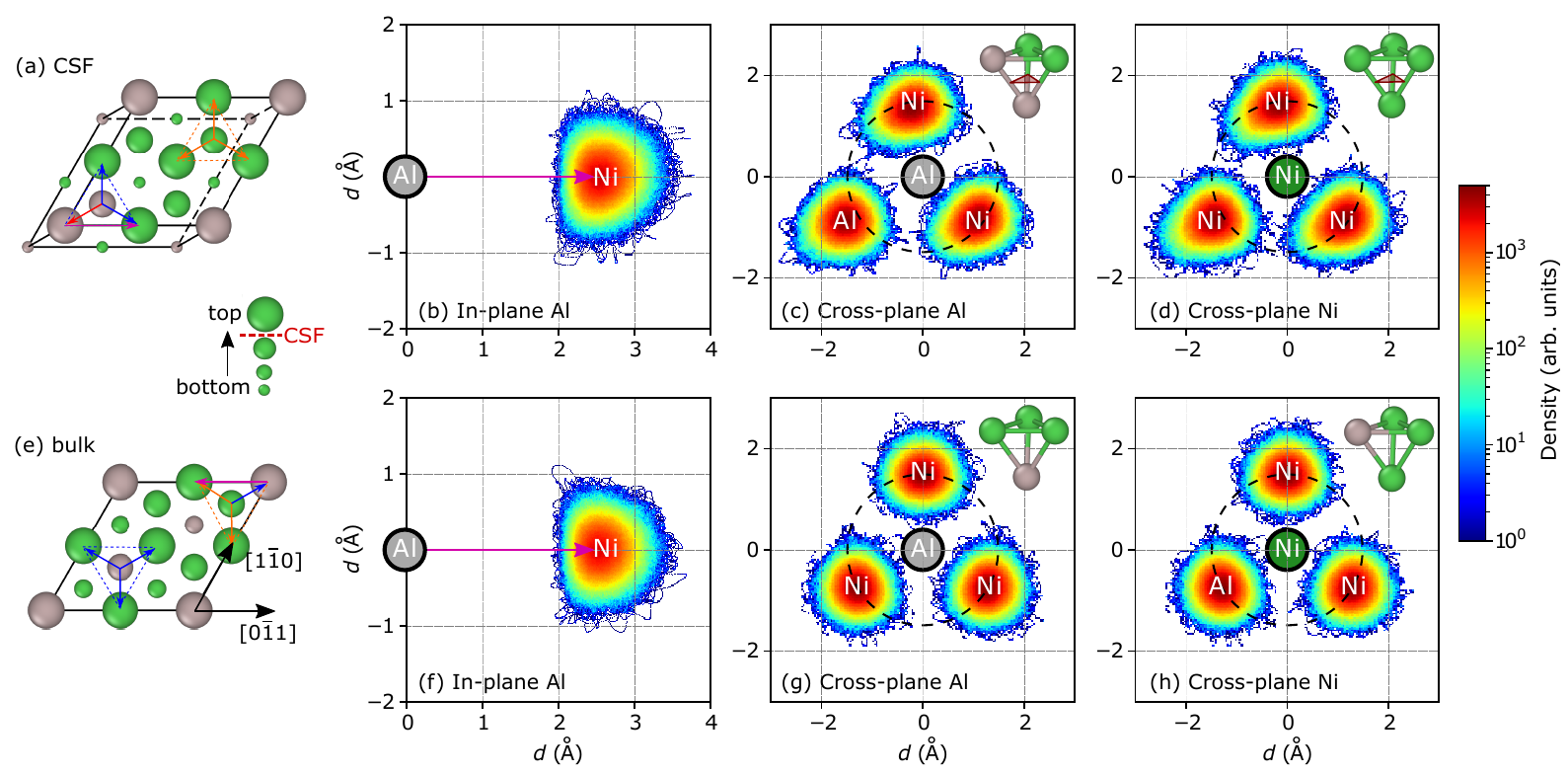}
\caption{Distributions of projected 1NN pairs for both the CSF and bulk Ni$_3$Al at a temperature of 1454\,K and at a lattice constant of 3.65~\AA\ \textcolor{black}{obtained from MD with the here optimized MTP}.
(a) and (e) 
Illustration of the local chemical environment for the CSF and bulk, respectively. Atoms are projected onto the (111) plane. Arrows indicate different types of atomic pairs, i.e., Al-Ni (magenta for in-plane and blue for cross-plane), Ni-Ni (orange) and Al-Al (red).
(b) to (d) Pair distributions in the CSF structure and (f) to (h) pair distributions in  bulk Ni$_3$Al.
Dashed black circles represent the 1NN distance at a lattice constant of 3.65~\AA~in bulk Ni$_3$Al.
The tetrahedrons inside the panels give the three-dimensional illustration of the given pair interactions (cf.~tetrahedrons in Figure~\ref{fig:CSF}).
}
\label{fig:VibDis}
\end{figure*}

Figure~\ref{fig:CSFE} presents the calculated temperature-dependent CSF Gibbs energy for Ni$_3$Al.
The black solid curve  in Figure~\ref{fig:CSFE}(a) shows the final CSF energy including all relevant thermal effects.
A decrease of the CSF energy with temperature from 300~K onward is observed. At the melting temperature of 1668\,K, the CSF energy has decreased by almost $50$ mJ/m$^2$ with respect to its room temperature value. All of the contributions are of a similar size and thus important to consider (Figure~\ref{fig:CSFE}(b)). They can be positive and act to destabilize the CSF (electronic and quasiharmonic) or negative and stabilize the CSF (anharmonic and spin fluctuations). 
Particularly at high temperatures, the explicit anharmonic vibrations show a strong impact on the CSF energy 
with a maximum decrease of 47 mJ/m$^2$ at the melting point. 
The spin fluctuations likewise provide a significant contribution to the CSF energy 
further decreasing it by up to 53 mJ/m$^2$.
The electronic free energy including the coupling to thermal vibrations increases with temperature up to 1200\,K by a similar magnitude as the quasiharmonic contribution.

The physical origin of anharmonicity in bulk materials~\cite{glensk2015understanding,srinivasan2023anharmonicity} and for point defects~\cite{Glensk2014PRX,gong2018Nivac} has been comprehensively understood from investigations of the local vibrational phase space.
It has been shown that anharmonic effects can be mostly captured by local pairwise interactions.
Here, in Figure~\ref{fig:VibDis}, we 
extended this analysis to planar defects and reveal the origin of the anharmonicity in the CSF energy by a comparison of the first nearest-neighbor (1NN) vibrational distribution for the CSF and bulk.
\textcolor{black}{To this end, we rely on our optimized machine-learning potentials, which provide the necessary accuracy at  minimal computational time.}
For the CSF structure, we focus on the local environments in the vicinity of the CSF plane where atomic arrangements feature tetrahedrons with different chemical decorations (see Figure~\ref{fig:CSF}).
Within these tetrahedrons, two types of 1NN atomic pairs can be distinguished, i.e., pairs within the basal (111) plane (``in-plane'') and pairs across the CSF plane (``cross-plane''). The distributions of one ``in-plane'' pair (Figure~\ref{fig:VibDis} (b)) and two different ``cross-plane'' pairs projected onto the (111) plane ((c) and (d)) are shown. The distributions for similar local environments in the bulk are shown in (f)--(h) for comparison. The corresponding 2D-projected geometrical relations are illustrated in (a) and (e).

The typical characteristic of anharmonic vibrations \cite{glensk2015understanding,srinivasan2023anharmonicity}, i.e., the breakdown of the distribution symmetry due to Pauli repulsion, is clearly observed in Figure~\ref{fig:VibDis} for both the CSF and the bulk, particularly for the projections of the ``in-plane'' pairs ((b) and (f)). 
This asymmetry is completely absent in the harmonic distributions (Figure~S3 in Supplementary Information).
The differences between the CSF and bulk distributions become prominent when the ``cross-plane'' pair interactions are compared for Al-centered pairs ((c) and (g)) and for Ni-centered pairs ((d) and (h)). While for bulk the centers of the distribution profiles locate at symmetrically equivalent positions (the equilibrium 1NN distances shown by the black dashed circles), the CSF specific pairs show remarkably elongated bond lengths. 
This feature can be understood from differences in the corresponding local environments.
With the presence of the CSF, chemically- and/or geometrically-distinguishable 1NN pairs are formed as compared to the bulk.
Specifically, the 1NN Al-Al pair is chemically unique to the CSF and it exhibits significant repulsive interactions
increasing its average bond length to about 2.84~\AA\ (Figure~\ref{fig:VibDis}(c)), 
 which is close to the 1NN distance of pure fcc Al (2.86~\AA~at 0~K).
Likewise, the geometrically distinct 1NN Ni-Ni pair, bottom left corner in Figure~\ref{fig:VibDis}(d), is also energetically unfavorable.
These repulsive interactions make the outward vibrations of the atom pairs more favorable, resulting in asymmetric oval-type distributions.
These remarkable differences in the vibrational distributions between CSF and bulk explain the observed large anharmonic contribution to the CSF energy.

\begin{figure}[t!]
\centering
\includegraphics[scale=0.7]{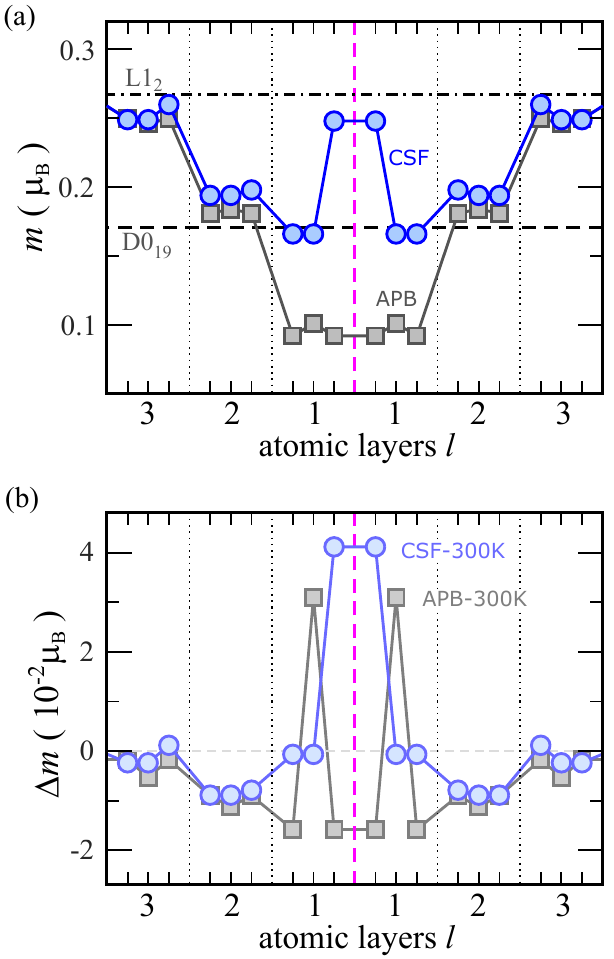}
\caption{(a) Magnetic moments of Ni-sites in the CSF and (111)APB structure from spin-polarized calculations at 0\,K.
Each atomic layer $l$ perpendicular to the (111) direction has three Ni atoms, two of which are equivalent.
(b) The difference of magnetic moments $\Delta m$ of Ni-sites between bulk and the defect structures for the PM state at 300\,K. The vertical magenta lines represent the position of planar defects.}
\label{fig:LSF}
\end{figure}

The change of the local chemical environments across the CSF plane, responsible for anharmonicity, also has a strong impact on the local magnetic moments of Ni and thus on the magnetic free energy.
For the FM case depicted in Figure~\ref{fig:LSF}(a), the presence of the CSF reduces the magnetic moments of most Ni atoms in the $l=1$ layer down to the values found for D0$_{19}$ Ni$_3$Al (an ordered hexagonal close-packed structure) as marked by the gray dashed line, due to the similarity in the local environment.
For the PM state, the difference in the magnetic moments $\Delta m$ between the defect phase  and bulk at 300~K is shown in Figure~\ref{fig:LSF}(b).
An enhancement of the magnetic moments with temperature at the CSF ($l=1$ layer) is observed, indicating a positive magnetic entropy contribution. 
\textcolor{black}{For completeness we note that the magnetic moments on the Al atoms are practically zero, since a large energy is required to induce a sizable magnetic moment due to the absence of \textit{d} states.}

In general, by comparing the CSF and the APB results, we observe qualitatively a similar impact of the planar defects and spin fluctuations on the local magnetic moments: A reduction of local magnetic moments near the defect plane in the FM state and an enhancement of the moments by spin fluctuations in the PM state at elevated temperatures. Quantitatively, in the PM state, the total difference in the enhanced magnetic moments for the CSF is larger than for the APB, suggesting a more positive magnetic entropy $\Delta S^{\mathrm{mag}}$ for the CSF than for the APB. Indeed, as temperature
increases, the $-T \Delta S^{\mathrm{mag}}$ term becomes more dominant and provides a more negative magnetic free energy, as reflected by the comparison of the spin-fluctuation contribution between CSF (green dots) and APB (green dashed line) in Figure~\ref{fig:CSFE}(b).

It should be noted that the predicted magnetic free energy from spin fluctuations is reliable at temperatures above 300\,K (well above the Curie temperature of 41.5\,K~\cite{de1969exchange}). 
For low temperatures approaching 0~K, the ferromagnetic approximation provides a reasonable estimation, as shown by the blue cross in Figure~\ref{fig:CSFE}(a). \textcolor{black}{Recent developments of magnetic machine-learning potentials \cite{brannvall2022machine,novikov2022magnetic} may help to improve the description around the Curie temperature in the future.}

Figure~\ref{fig:CSFE}(a) also contains the available experimental data  for comparison with the calculated CSF energy.
Unfortunately, in the temperature range above 500\,K, there is a lack of experimental data.
Considering the experimental uncertainty, 
\textcolor{black}{the value at 485\,K~\cite{hemker1993measurements} agrees reasonably well with the theoretical prediction, while the other two values at room temperature show a larger discrepancy from the present simulation result.}
It is important to note that the calculated CSF energy corresponds to the formation energy of a fully equilibrated CSF,
which is hardly achievable with samples deformed at low temperatures due to the complicated stress and strain field experienced by the CSF ribbon. \textcolor{black}{Likewise, it should be noted that different exchange-correlation functionals may result in different CSF energies, in particular, in an overall shift of the whole curve. Temperature dependencies are typically well described by the utilized GGA-PBE functional \cite{forslund2023thermodynamic,jung2023high}. Importantly, the computed thermodynamic properties of Ni$_3$Al agree well with experiment \cite{xu2023APBs}.}

The predicted temperature dependence of the CSF energy sheds new light on the understanding of the cross-slip process and the YSA.
\textcolor{black}{
A lower CSF energy results in a larger CSF ribbon and thus a higher enthalpy for the cross-slip from a (111) plane to a (100) plane.
To explain the YSA, several models have applied \textit{temperature independent} formation energies of the planar defects~\cite{hirsch1992model,caillard1997role,choi2007modelling,demura2007athermal}.
The decreasing CSF energy obtained in this work demonstrates an increasing energy barrier with temperature for the cross-slip process, hindering the occurrence of cross-slip events at elevated temperatures.
As a consequence, previous models based on the approximation of a temperature independent CSF energy need to be refined.
}

In summary, we have presented an accurate prediction of the temperature dependent CSF Gibbs energy in Ni$_3$Al up to the melting point by employing a recently proposed \textit{ab initio} framework that simultaneously takes all relevant thermal effects into account.
\textcolor{black}{The CSF energy decreases with temperature, due to a strong contribution from spin fluctuations and anharmonic vibrations. 
By applying a new analysis focused on the vibrational space of atomic pairs, we have found a strong influence of the CSF geometry on the anharmonic distributions of the cross-plane atomic pairs.
The calculated decreasing CSF energy indicates a higher energy barrier for the cross-slip process at higher temperatures, suggesting a necessity to modify the existing constitutive models.
The predicted energies for CSF and two types of APBs~\cite{xu2023APBs} are expected to advance the interpretation of the yielding behavior of intermetallic materials, particularly the YSA, and facilitate the design of the next-generation superalloys.
}



\section*{Data Availability}
The authors declare that all data supporting the findings are available from the corresponding author upon reasonable request. 

\section*{Competing interests}
The authors declare no competing interests.

\section*{Acknowledgements}
This work has been funded by the Deutsche Forschungsgemeinschaft (DFG, German Research Foundation) under the Germany's Excellence Strategy - EXC 2075 – 390740016. 
We acknowledge the support by the Stuttgart Center for Simulation Science (SimTech).
We acknowledge the support by the state of Baden-Württemberg through bwHPC and the German Research Foundation (DFG) through grant no INST 40/575-1 FUGG (JUSTUS 2 cluster).
The authors gratefully acknowledge the Gauss Centre for Supercomputing e.V.
(www.gauss-centre.eu) for funding this project by providing computing time
on the GCS Supercomputer HAWK at H{\"o}chstleistungsrechenzentrum Stuttgart
(www.hlrs.de).
B.G. acknowledges funding from the European Research Council (ERC) under the European Unions Horizon 2020 research and innovation programme (Grant Agreement No.~865855).
A.R. acknowledges the support by the Austrian Federal Government (in particular from Bundesministerium f{\"u}r Verkehr, Innovation und Technologie and Bundesministerium f{\"u}r Wirtschaft, Familie und Jugend) represented by {\"O}sterreichische   Forschungsförderungsgesellschaft mbH and the Styrian and the Tyrolean Provincial Government, represented by {\"O}sterreichische      Forschungsförderungsgesellschaft mbH and Standortagentur Tirol within the framework of the COMET Funding Programme. Calculations have been partly done using NSC (Link\"oping) and PDC (Stockholm) resources provided by the Swedish National Infrastructure for Computing (SNIC).




\section*{Supplementary material}

# Supplementary Information to "Accurate temperature-dependent complex-stacking-fault Gibbs energy in Ni$_3$Al"

Xiang Xu[1,2], Xi Zhang[2] [*], Andrei Ruban[3,4], Siegfried Schmauder[1], and Blazej Grabowski[2]

[1]*Institute for Materials Testing, Materials Science and Strength of Materials, University of Stuttgart, Pfaffenwaldring 32, 70569 Stuttgart, Germany*
[2]*Institute for Materials Science, University of Stuttgart, Pfaffenwaldring 55, 70569 Stuttgart, Germany*
[3]*KTH Royal Institute of Technology, SE-100 44 Stockholm, Sweden*
[4]*Materials Center Leoben Forschung GmbH, A-8700 Leoben, Austria*
[*]*Corresponding author: xi.zhang@imw.uni-stuttgart.de*

## Supp. 1    eDOS

Figure S1 and S2 show the partial electronic density of states of Ni atoms in bulk Ni$_3$Al and in the CSF structure (the tilted cell in Figure 2 in the main text) at zero-Kelvin. The influence of inserting a CSF plane decays when departing from the CSF plane and almost vanishes when $l = 3$, confirming that a model with six atomic layers is sufficient to exclude the interaction between the CSF and its periodic images.

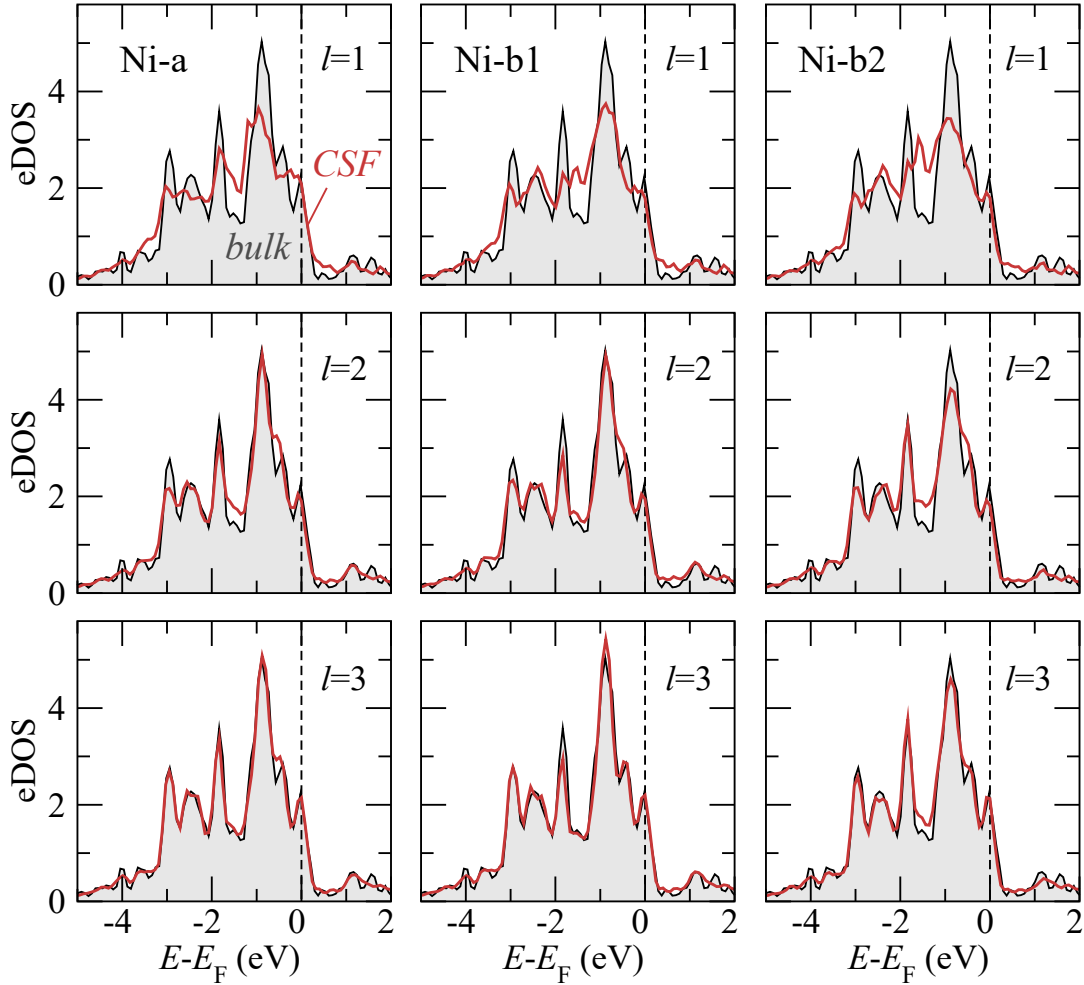

Figure S1: Partial electronic density of states (eDOS) for Ni atoms in the CSF structure and bulk Ni$_3$Al from spin-unpolarized calculations. $E_\text{F}$ refers to the Fermi energy. The value of $l$ gives the distance to the CSF in terms of atomic layers. Each atomic layer has three Ni atoms labeled as "Ni-a", "Ni-b1" and "Ni-b2" ("Ni-b1" and "Ni-b2" are equivalent).

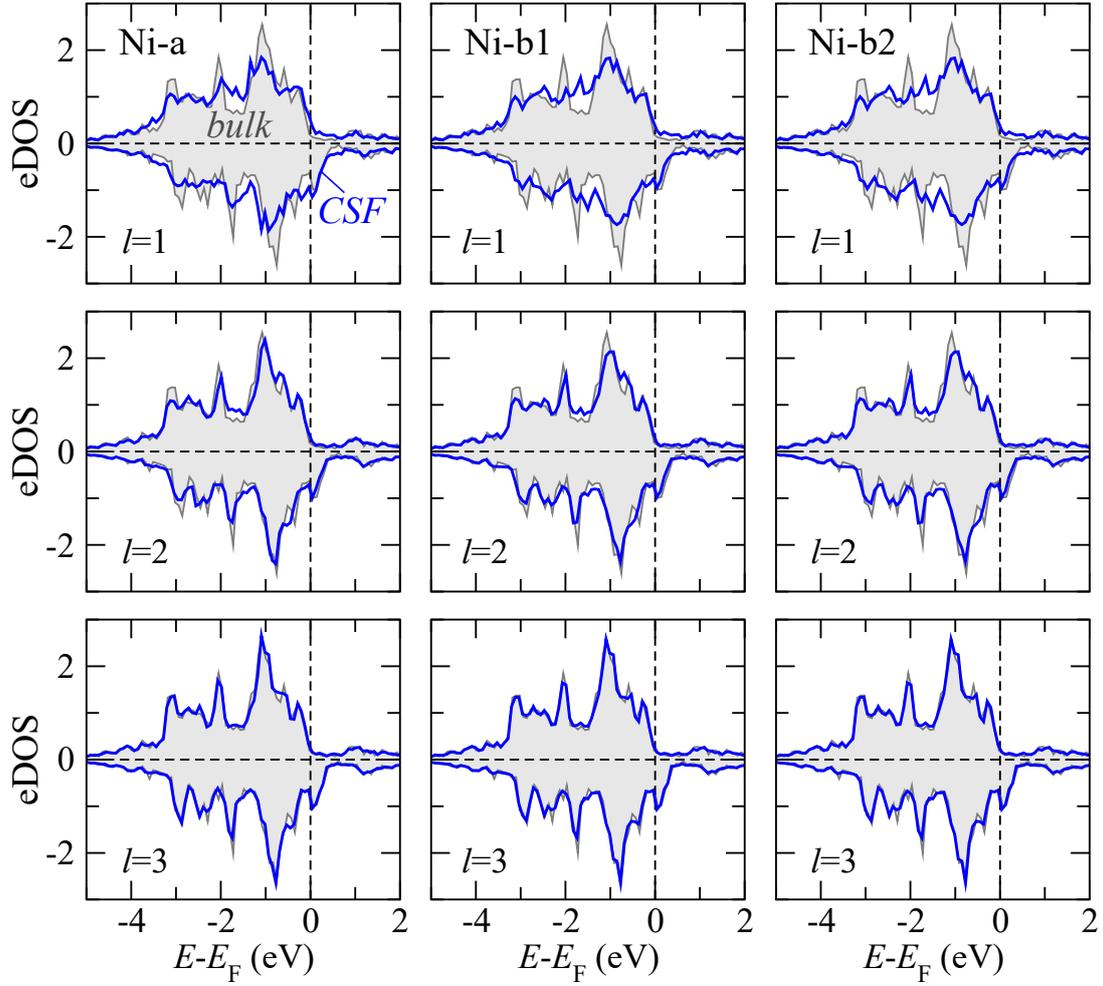

Figure S2: Partial electronic density of states (eDOS) for Ni atoms in the CSF structure and bulk $Ni_3Al$ from spin-polarized calculations.

## Supp. 2 Distribution of Pair Interactions

Distributions of several millions of atomic pairs are collected from MD simulations with using the Langevin dynamics with the fitted MTP. Here, the trained MTP was utilized, which ensures the accuracy and efficiency of the vibrational phase space. When including the anharmonicity, the first nearest-neighbor (1NN) pair distribution shows a typical wall-like planar edge [1,2], while the quasiharmonic distribution is symmetric, as shown in Figure S3. It is interesting that the in-plane 1NN pair interactions demonstrate similar distributions for both bulk and CSF structures. The anharmonic term in Equation (2) in the main text can be characterized by the difference between the full-vibration distribution and the QH-vibration distribution.

On the other hand, the cross-plane 1NN distributions present an evident deviation between bulk and CSF structures, as shown in Figure S4. This difference is assumed to be the reason for the substantial anharmonic contribution to the CSF energy at higher temperatures.

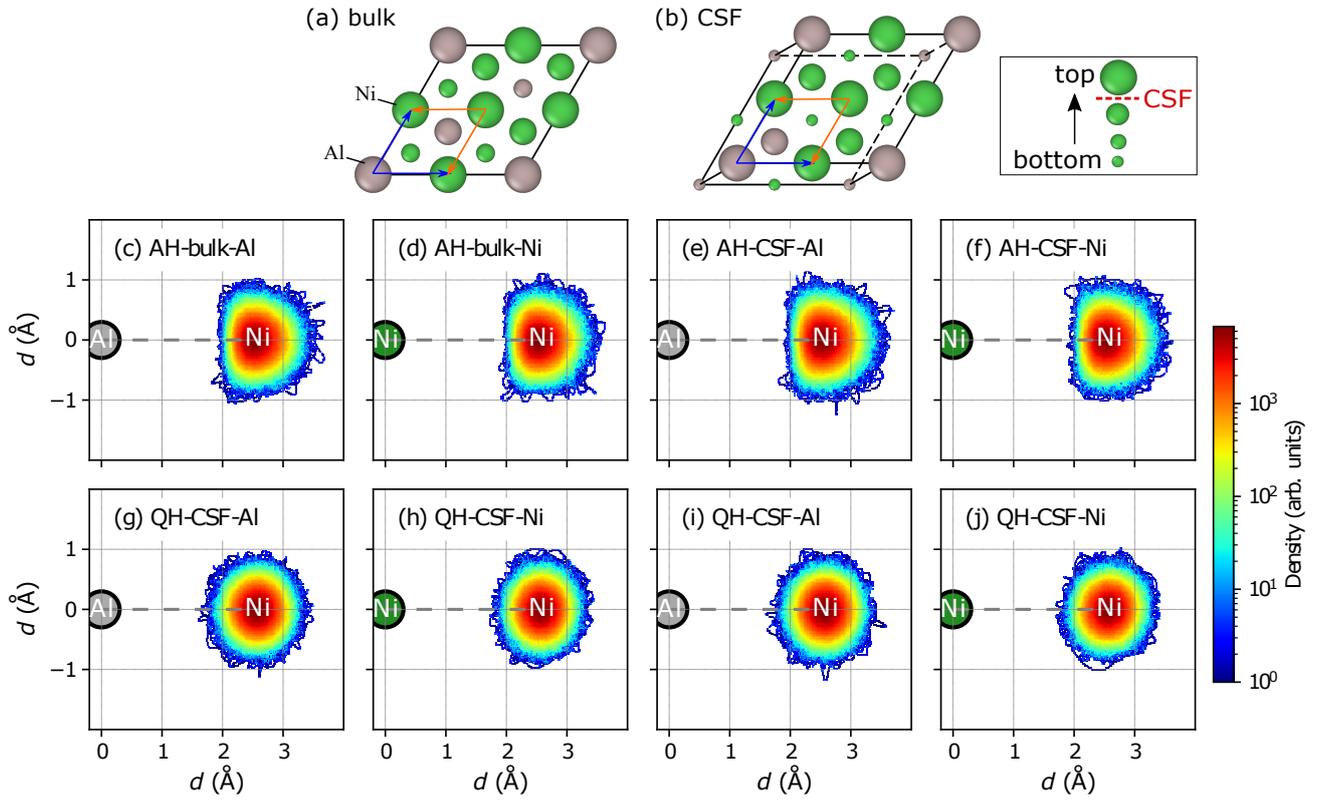

Figure S3: In-plane distributions in Ni$_3$Al and CSF structures at 1454 K and at 3.65 Å. (a) and (b) are the projection on (111) plane of bulk and CSF structure. Blue and magenta arrows indicate the Al-Ni and Ni-Ni pairs, respectively. (c) to (d) are the distribution considering full vibrations, while (g) to (j) are only from quasiharmonic force constants.

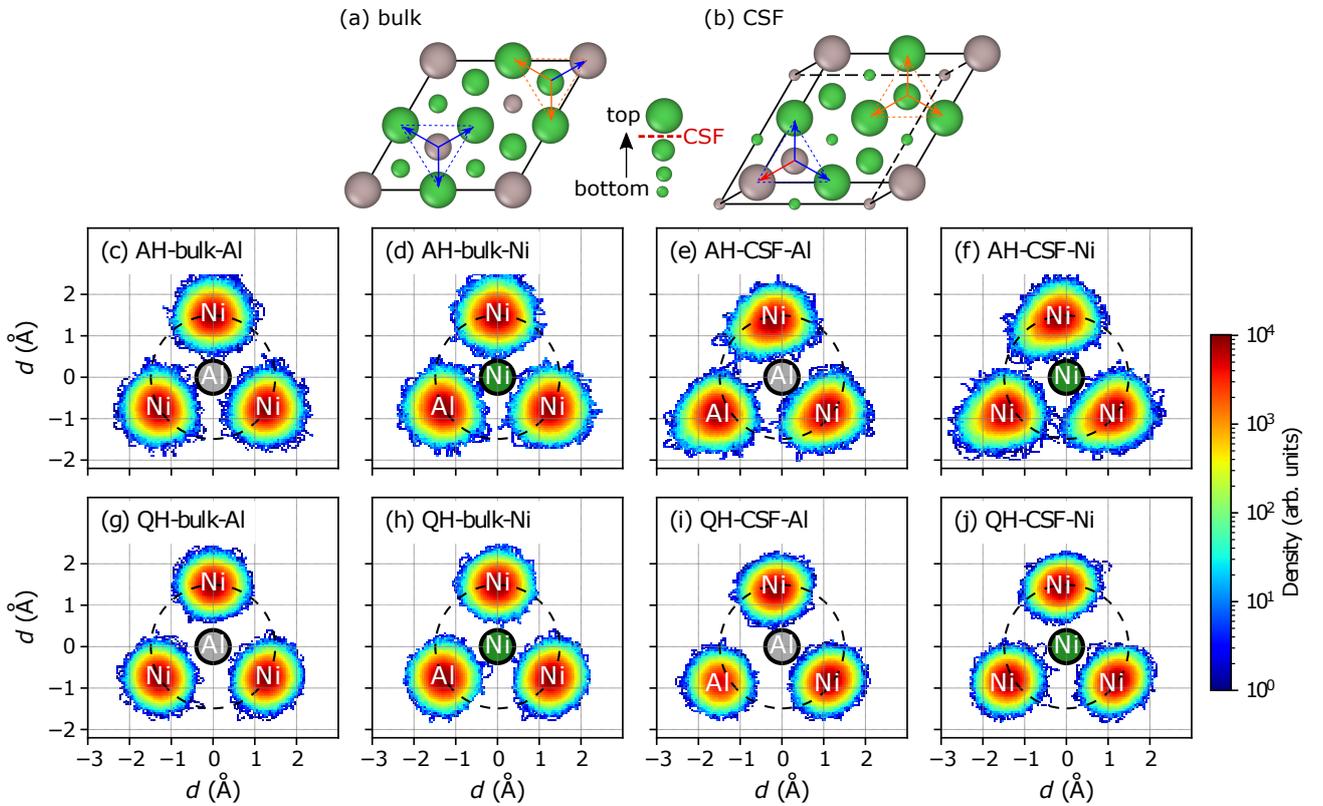

Figure S4: Cross-plane distributions in Ni$_3$Al and CSF structures at 1454 K and at the corresponding equilibrium lattice constant of 3.65 Å. (a) (d) The atomic positions projected onto the (111) plane. Blue, orange and red arrows indicate the Al-Ni, Ni-Ni and Al-Al pairs, respectively. (c) to (d) are the distribution considering full vibrations, while (g) to (j) are only from quasiharmonic force constants. Dashed black circles represent the 1NN distance at a lattice constant of 3.65 Å in bulk Ni$_3$Al.

## Supp. 3 Collection of CSF energies

Table S1: Collection of CSF energies in mJ/m$^2$. Experimental values marked with "†" are shown in Figure 3 in the main text. RT: room temperature. EAM: Embedded Atom Model; TB-LMTO: Tight Binding Linear Muffin-Tin Orbitals; FLMTO-PN: Full-Potential Linear Muffin-Tin Orbital within the Peierls–Nabarro model.

| Experiments | | | |
|---|---|---|---|
| Reference | Composition | Energies (mJ/m$^2$) | $T$ (K) |
| Hemker1993 [3] | Ni$_{76}$Al$_{24}$ | $206 \pm 30$ | 485 |
| Karnthaler1996 [4] | Ni$_{78}$Al$_{22}$ | $235 \pm 40$ | RT |
| Kruml2002 [5] | Ni$_{76}$Al$_{24}$ | $206 \pm 27$ | RT |
| Kruml2002 [5] | Ni$_{75}$Al$_{25}$ | $236 \pm 29$ | RT |
| Kruml2002 [5] | Ni$_{74}$Al$_{26}$ | $277 \pm 49$ | RT |
| Baluc1991 [6] | Ni$_{74.3}$Al$_{24.7}$Ta$_{1.0}$ | $300 \pm 40$ | RT |
| Hemker1993 [3] | Ni$_{75.5}$Al$_{23.8}$B$_{0.7}$ | $335 \pm 60$ | 485 |

| Theoretical calculations at 0 K for Ni$_3$Al | | | |
|---|---|---|---|
| Reference | Method | Energies (mJ/m$^2$) | Magnetic state |
| *present work* | DFT, PAW-PBE | 217 | FM |
| *present work* | DFT, PAW-PBE | 171 | NM |
| Rosengaard1994 [7] | DFT, TB-LMTO | 308 | FM |
| Mryasov2002 [8] | DFT, FLMTO-PN | 225 | |
| Yu2012 [9] | DFT, PAW-PBE | 202 | FM |
| Liu2015 [10] | DFT, PAW-PBE | 208 | FM |
| Rao2018 [11] | DFT, PAW-PW91 | 187 | FM |
| Mishin2004 [12] | EAM | 202 | |
| Chen1989 [13] | EAM | 120 | |
| Du2012 [14] | EAM | 218 | |



\end{document}